\documentclass[11pt]{article}
\usepackage{graphicx}
\usepackage[margin=1.25in]{geometry}
\usepackage[usenames,dvipsnames]{color}
\usepackage{url}
\usepackage{wrapfig,rotating}
\usepackage[colorlinks = true,
            linkcolor = blue,
            urlcolor  = blue,
            citecolor = blue,
            anchorcolor = blue]{hyperref}

\usepackage{amssymb, booktabs, color, xcolor, etaremune, eurosym, fancyhdr, graphicx, lastpage, layout, lipsum, multicol, multirow, paralist, pbox, pdfpages, rotating, sidecap, titlesec, units, url, verbatim, wrapfig, xspace, tabularx,cite, sidecap, inputenc}

\newcommand\pubnumber{Transcendental Preprint }
\newcommand\pubdate{\today}

\def\Title#1{\begin{center} {\LARGE #1 } \end{center}}
\def\Author#1{\begin{center}{ \sc #1} \end{center}}
\def\Address#1{\begin{center}{ \it #1} \end{center}}

\newcommand\pubblock{\rightline{\begin{tabular}{l} \pubnumber\\
         \pubdate \end{tabular}}}
\newenvironment{Abstract}{\begin{quotation} \begin{center}
                       ABSTRACT
     \end{center}\bigskip  }{\end{quotation}}

\def\eg{{\it e.g.}}

\def\beq{\begin{equation}}
\def\eeq#1{\label{#1}\end{equation}}
\def\eeqn{\end{equation}}

\newenvironment{Eqnarray}%
   {\arraycolsep 0.14em\begin{eqnarray}}{\end{eqnarray}}
\def\beqa{\begin{Eqnarray}}
\def\eeqa#1{\label{#1}\end{Eqnarray}}
\def\eeqan{\end{Eqnarray}}

\let\bar=\overbar

\def\lsim{\mathrel{\raise.3ex\hbox{$<$\kern-.75em\lower1ex\hbox{$\sim$}}}}
\def\gsim{\mathrel{\raise.3ex\hbox{$>$\kern-.75em\lower1ex\hbox{$\sim$}}}}

\def\del{\partial}
\def\Dslash{\not{\hbox{\kern-4pt $D$}}}
\def\dslash{\not{\hbox{\kern-2pt $\del$}}}
\def\pslash{\not{\hbox{\kern-2pt $p$}}}
\def\ETmiss{\not{\hbox{\kern-4pt $E$}}_T}

\def\Dlr{\mathrel{\raise1.5ex\hbox{$\leftrightarrow$\kern-1em\lower1.5ex\hbox{$D$}}}}

\def\MSB{{\bar{M \kern -2pt S}}}
\def\msb{{\bar{\scriptsize M \kern -1pt S}}}

\def\drb{{\bar{\scriptsize D \kern -1pt R}}}

\newcommand\snowmass{\begin{center}\rule[-0.2in]{\hsize}{0.01in}\\\rule{\hsize}{0.01in}\\
\vskip 0.1in Submitted to the  Proceedings of the US Community Study\\ 
on the Future of Particle Physics (Snowmass 2021)\\ 
\rule{\hsize}{0.01in}\\\rule[+0.2in]{\hsize}{0.01in} \end{center}}

\begin{document}

\pubblock

\Title{Light-weight and highly thermally conductive support structures for future tracking detectors}

\bigskip 

\Author{E.\ Anderssen$^{1}$, A.\ Jung$^{2}$, S.\ Karmarkar$^{2}$, A.\ Koshy$^{2}$}

\medskip

\Address{   $^{1}$ LBL, USA\\
            $^{2}$ Purdue University, USA}

\medskip

 \begin{Abstract}
\noindent Detector mechanics can play a significant role in a detector's performance, improvements typically require in-depth study of total mass, novel ways to reduce the total mass, as well as more integrated design concepts to save on material budgets and optimize performance. Particle detectors at future colliders rely on ever more precise charged particle tracking devices, which are supported by structures manufactured from composite materials. This article lays out engineering techniques able to solve challenges related to the design and manufacturing of future support structures. Examples of current efforts at Purdue University related to the high-luminosity upgrade of the CMS detector are provided to demonstrate the prospects of suggested approaches for detectors at new colliders: a future circular collider or a muon collider. Detectors at electron-positron machines have significantly smaller material budgets and require targeted concepts. 
\end{Abstract}

\snowmass

\def\thefootnote{\fnsymbol{footnote}}
\setcounter{footnote}{0}

\section{Introduction}
\label{toc:Intro}
Future particle colliders, such as the high luminosity Large Hadron Collider (HL-LHC), the Future Circular Collider (FCC-hh) or the muon collider (MuC) will collide particles at unprecedented rates to search for new physics and make high precision measurements to challenge the standard model. The increase in granularity of the detectors and the background rates pose high demands for the materials of charged particle tracking detector support structures. State-of-the-art silicon tracking detectors are typically large systems assembled from thousands of individual silicon modules each consisting of silicon sensors (pixel, strip or pixel-strip), read-out chips, as well as associated on-detector electronics. These highly-integrated electronic circuits produce heat that has to be efficiently removed, \eg \ through the support structure, to keep the electronics at the optimal operation temperature and prevent thermal runaway of the sensors. 
Tracking detectors at current (and future) colliders are typically exposed to a high-radiation environment where carbon fiber composite materials are employed to design and build the mechanical support structures of silicon detectors because of their high thermal conductivity, strength to mass ratio and radiation tolerance. Silicon detectors facilitate precision position measurements and hence, require mechanical structures to be accurately assembled and precisely manufactured to allow ideal coverage of the interaction point\footnote{The achievable module alignment resolution is of order 1 micrometer for the inner pixel detector and of order 3 micrometer for the outer strip tracker of the current CMS detector\cite{Run2TrackerPerformance}.}.
The future of the manufacturing process of these structures should be revisited to optimize support mass towards minimal weight systems, also potentially integrating services into structures, all in order to boost the detector performance in terms of resolution and readout rate capability. Of course, at the same time we need to pay attention to deflections that arise from dynamical loads during installation but also due to gravitational sag and thermal deflections. If not modeled accurately any of these can prevent the integration of the larger detector systems or reduce the expected precision of the detector. 
These deflections can nowadays be predicted fairly accurately by finite element analysis (FEA), such that they can be compensated and ultimately mitigated in the manufacturing of a final mechanical support structure. The particle physics community invested in detector mechanics has always harvested and utilized the rapid progress in composite manufacturing in industry and adopt or partner to have the best solutions for particle physics detector support structures, and will continue to do so into the future.

\section{Towards future tracking detector support structures}
\label{toc:AimOfArticle}
In this article we highlight a few of the current challenges related to the design of tracking detector support structures. By exploiting the experience gained from major responsibilities related to the HL-LHC upgrade of the CMS detector \cite{PhaseII_TDR}, namely design and manufacturing of multiple large mechanical support structures for the Inner and Outer Silicon Tracker of CMS. We briefly summarize challenges in state-of-the-art support structures in Section \ref{toc:StateOfTheArt}, followed by an introduction into how to design support structures in Section \ref{toc:compositeStructuresIntro} using latest techniques in composite engineering towards multi-functional support structures harvesting gains of machine learning for optimizing the detector mass. 
\begin{SCfigure}
\includegraphics[trim=0cm 0cm 0cm 1cm,width=0.6\textwidth,angle=0]{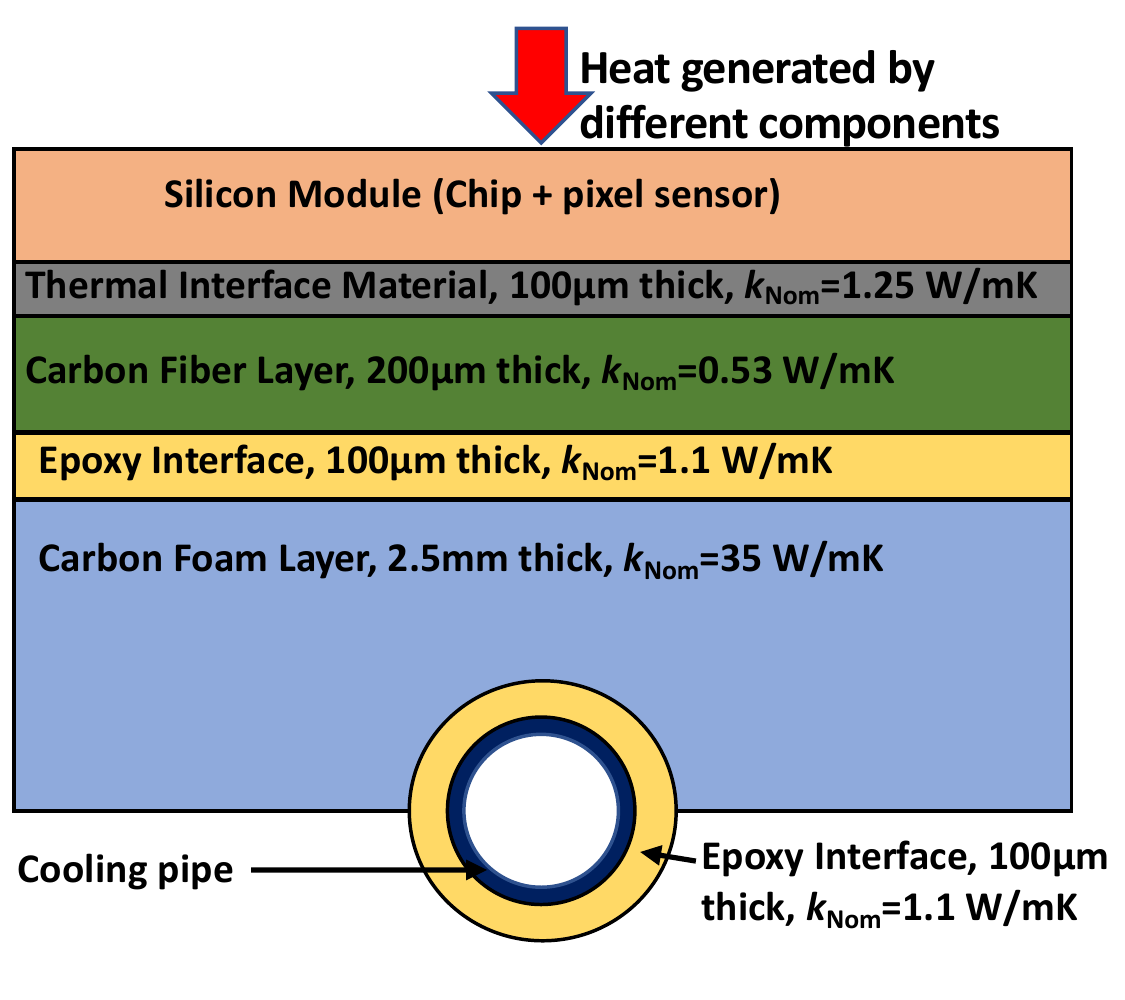}

\caption{Schematic view of the cross-section of a representative multi-layer support structure inspired by the CMS Inner Tracker half-disc support structure. The thicknesses and the typical nominal thermal conductivities of the chosen materials are provided as well, and the entire structure is symmetric around the cooling pipe but suppressed here for simplicity.}
\label{fig:module_schematic} 
\end{SCfigure}

Ongoing detector R\&D activities at Purdue University are towards more integrated cooling and mechanical support structures, which ultimately could be manufactured in a bottom-to-top approach using latest techniques such as 3D printing. Novel techniques, materials and other design and manufacturing solutions provide an avenue to solve challenges related to increasingly more complex tracking detectors at future colliders.

\section{State-of-the-art support structures}
\label{toc:StateOfTheArt}
Silicon tracking devices are employed in many particle detectors today and are foreseen to be heavily used in future detectors to measure precisely the trajectories of charged particles in high energy and nuclear physics. These devices are the backbone for a huge variety of precision measurements, which all rely on near 100\% efficient and very stable detector operating conditions. The large amount of heat generated in these low heat capacity sensors and their associated on-module electronics can damage them if the heat is not removed efficiently. This task becomes more demanding for irradiated sensors, with their increased leakage current, especially towards the end of the respective operation of a detector.\\
Among the various cooling mechanisms, natural convection and radiative cooling are insufficient for cooling of high-rate large silicon detector systems because of their limited heat removal capacity. Typically, the heat generated by the silicon modules is removed to a local heat sink or a cooling pipe by means of thermal conduction through an optimized support structure~\cite{PhaseII_TDR, PhaseII_TDR_ATLAS}. The temperature of the cooling pipe is kept stable independent of the heat generated by the sensors by circulating coolants like $ \mathrm{CO_{2}} $, which are maintained in a two-phase state with liquid and gas in equilibrium.\\

\begin{figure}[ht]

\center
\includegraphics[trim=0cm 0cm 0cm 0cm,width=\textwidth]{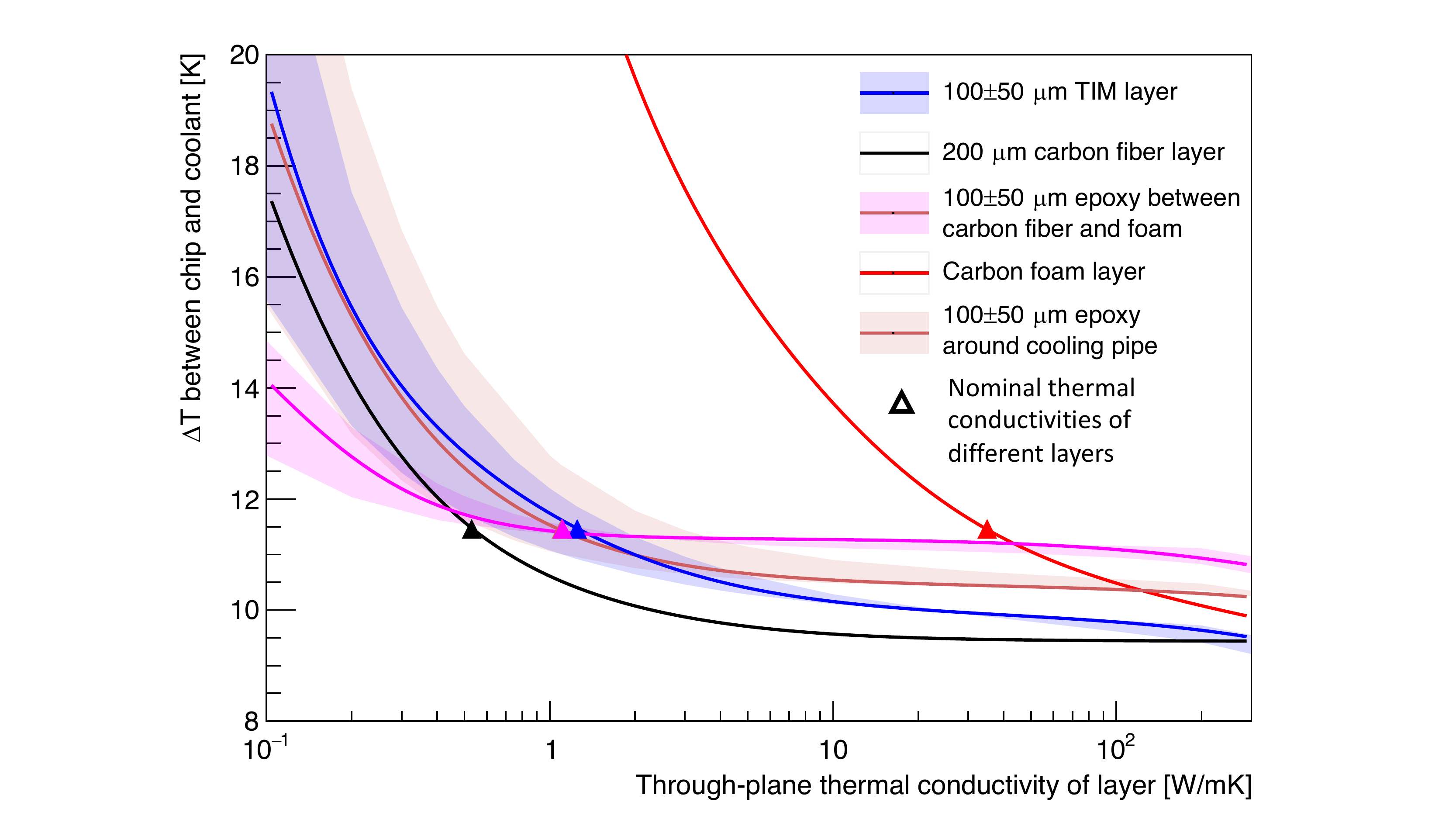}

\caption{The temperature difference between the hottest point on the chip and the coolant as a function of the thermal conductivities of different layers of the hypothetical support structure shown in Figure \ref{fig:module_schematic}. FEA simulation results are used to determine the functional form of the temperature dependence of the various materials. The shaded bands represent the impact of a $\pm 50 \mu m$ thickness variation for the various epoxy interfaces. More details in the text.}
\label{fig:TC_motivation} 
\end{figure}

The materials used for manufacturing mechanical support structures of the silicon tracking detector need to have high stiffness and excellent thermal conductivity (TC). Carbon fiber composite materials meet most of these requirements but have low thermal conductivites perpendicular to the carbon fiber direction. A schematic multi-layer ``stack" of materials is sketched in Figure \ref{fig:module_schematic}, with the cooling pipe at the bottom and the silicon device at the top. The entire structure is symmetric around the cooling pipe, but to ease calculations we only focus on the top half structure. The structure consists of a variety of materials, namely a representative silicon module as heat source, carbon fiber laminate, carbon foam, cooling pipe and the various thermal interface materials (TIM) between these layers. This structure is modeled in a finite element analysis (FEA) to study and identify the impact of individual layers on the overall thermal performance of the structure \cite{Koshy}. The conditions and input parameters of the FEA are as follows: the ambient temperature is set to T= 22$^{0}$C, a convective coefficient of $h=5\mathrm{W/m^{2}K}$ corresponding to steady air is employed to model heat exchange via convection, the cooling pipe is kept at T= -36$^{0}$C with a convective coefficient of $h=5000\mathrm{W/m^{2}K}$. Many independent FEA runs are used to determine the functional form of the thermal performance of this structure. Figure \ref{fig:TC_motivation} shows the temperature difference, $\Delta T$, of the silicon device with respect to coolant temperature as a function of the thermal conductivity of the various material and interface layers. In order to study the impact of the thermal conductivity of a specific layer on the chip temperature, the thermal conductivity of that layer is varied whilst keeping the TCs of all other layers at their nominal values. The shaded bands for the TIM layer and epoxy layer surrounding the cooling pipe are determined by varying the nominal $\pm100\mu m$ thickness of these layers by $\pm50\mu m$. 

We note a steep rise in this temperature difference around the theoretically expected value of carbon fiber through-plane TC that was calculated using Mechanics of Structural Genome (MSG) techniques~\cite{msg}. Given how close this nominal TC is to the rising edge of the functional form, small deviations in the through-plane TC result in a significant increase of the observed $\Delta T$. Of course the impact of other materials on the thermal performance also starts to be significant once a certain low TC of a specific material is chosen, but the currently available material solutions for these layers have nominal TCs that are more far away from the start of the temperature rise. This study demonstrates, generally, that the through-plane TC of carbon fiber will be critical for heat extraction from the silicon module. 

%
%
\section{Composite Structure Design for HL-LHC}
\label{toc:compositeStructuresIntro}

The ``traditional" approach to detector mechanics design starts with the design of silicon detectors that cover the active sensing area for a detector surrounding the interaction point including the electronics, power and readout design. At that stage of the design the support structures needed to precisely mount silicon modules, all services, as well as cooling is a bare bone design able to full-fill the task but at times and in the details lacking a sense for the realities of engineering constraints. The electronics, readout, and cooling systems take precedence over the composite support structure, though the entire silicon detector is constrained by the deflection requirements and thermal performance of the detector structure. Throughout the past and even into the current HL-LHC detector upgrades \cite{PhaseII_TDR_ATLAS,PhaseII_TDR} the iterative design process for the composite support structure is often a result of the cooling requirements of the detector. The mass of the support structure is a barrier to better detectors and thus, for future detectors like the FCC or MuC it will be the bottle neck that needs to be addressed. Improvements in structural design methodology for future detectors are a necessity to pave the way for better thermal management solutions integrated into the composite structures. We highlight latest techniques in composite engineering harvesting the power of machine learning to optimize the detector support mass. We utilize our ongoing efforts related to the HL-LHC CMS detector upgrade projects at Purdue University to highlight future directions in detector support mechanics.    

\subsection{Composite Structure Design Methodology}
\begin{figure}
    \centering
    \includegraphics[width=\textwidth,height=\textheight,keepaspectratio]{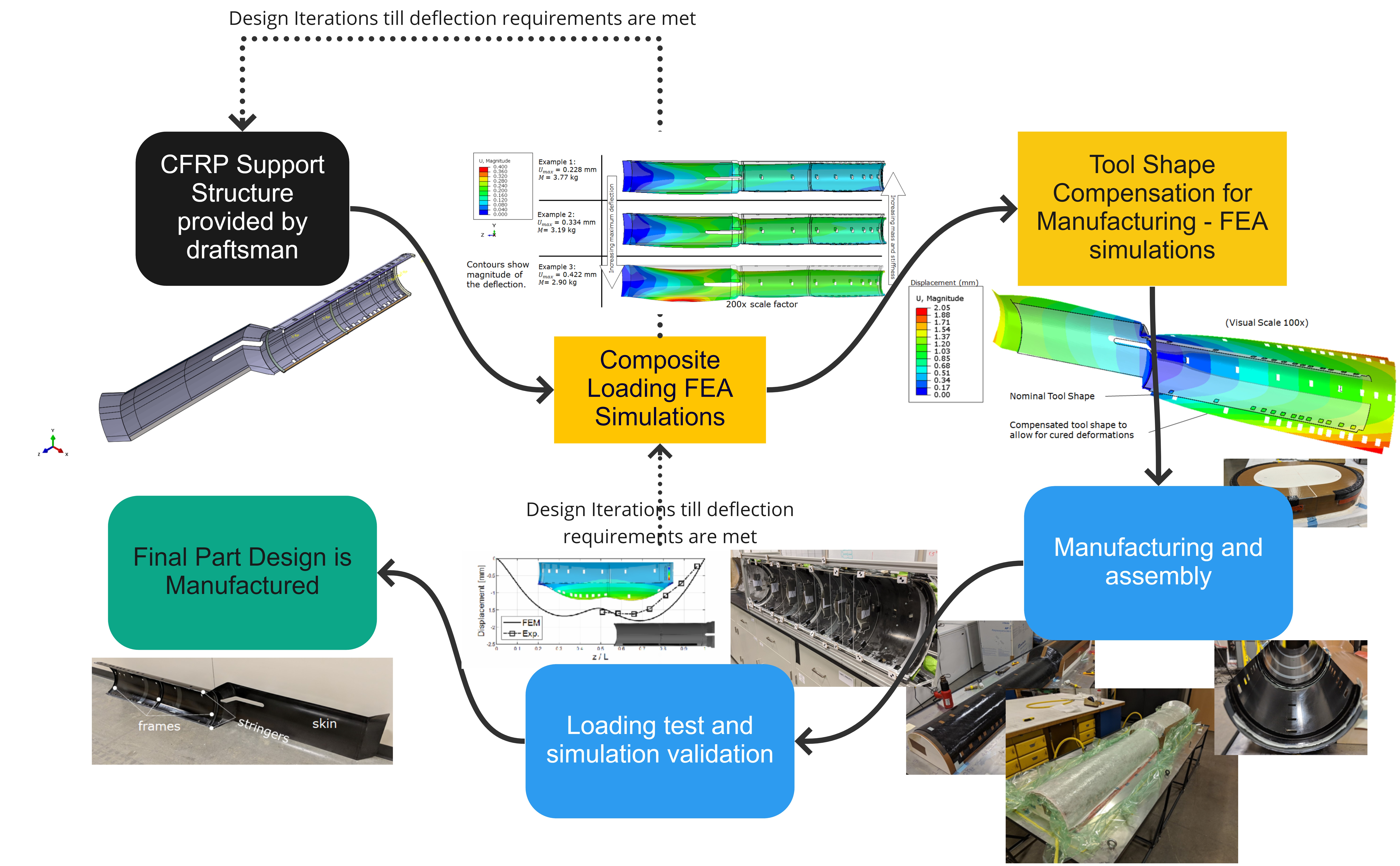}
    \caption{Sketch of the composite support structure design methodology followed in the context of CMS detector upgrades.}
    \label{fig:Overview1}
\end{figure}

The structural design process illustrated in Figure \ref{fig:Overview1} is used for the HL-LHC CMS support structures. This process involves end-to-end process and performance simulations for the carbon fiber composite structures. The integration of process as well as performance simulations for composite structures enables the design and manufacturing of detector support structures to the required tolerances which are typically on the scale of one to few millimeters for a part that is about 5 meters long. To enable such performance simulations a high fidelity model needs to be developed. The composite loading finite element simulations consist of the loading and boundary conditions that mimic the detector design considerations. The thermal simulations consider the temperature gradients on the part and are essential and critical due to the anisotropic nature of composite structures. 

\begin{figure}
    \centering
    \includegraphics[width=\textwidth,height=\textheight,keepaspectratio]{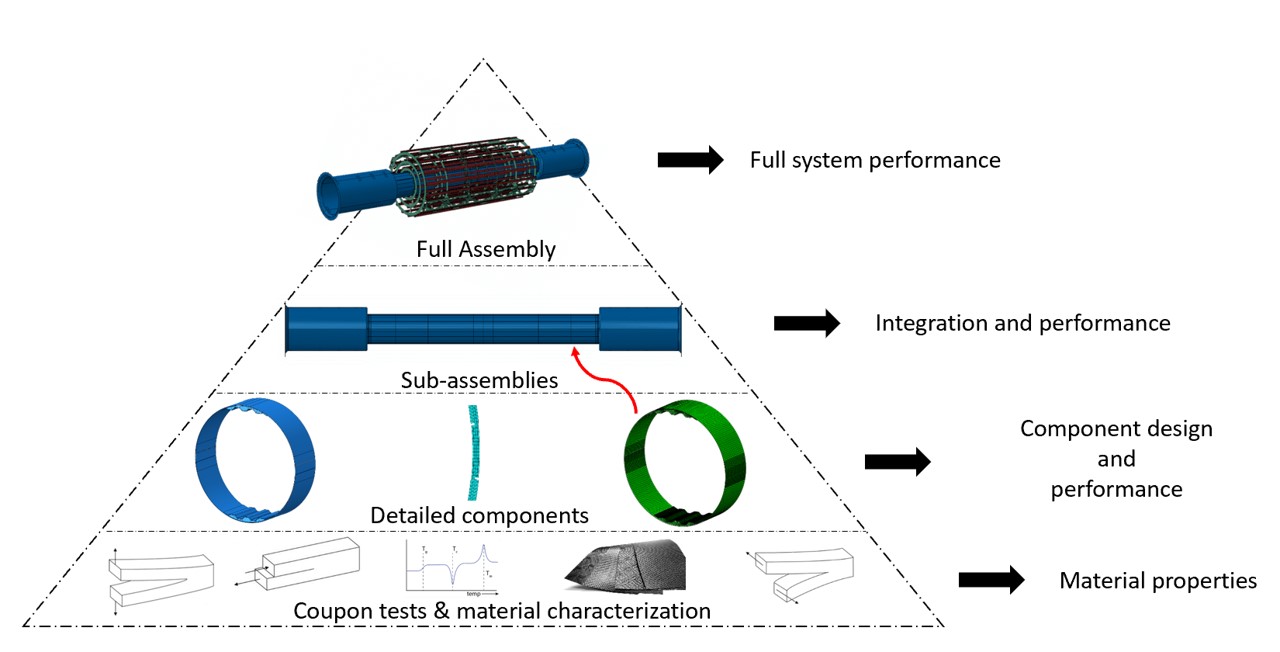}
    \caption{Working toward a higher-fidelity assembly model \cite{JustinHicksFTDM2021}
    }
    \label{fig:pyramid1}
\end{figure}

Any carbon fiber composite support structure typically consists of multiple layers of individual plies of carbon fiber, since it is typically delivered as single-ply pre-preg material. The ply stacking sequence and orientation is a vital design criteria to explore the mass-to-stiffness ratio optimization, which can be modeled by a mechanical FEA. This FEA simulation and design starts after a CAD model has been created based on the physics detector requirements and positioning of the silicon detectors. Given the millimeter tolerances it is important to get the loading conditions of the mounting of detectors, cooling pipes, readout system and power electronics accurately into the FEAs. Furthermore, the thermal gradients due to the differential cooling of the detector volume is an essential input to these FEA simulations since the deformations are non-negligible. The precision of the desired structure highly depends on getting the simulation parameters accurate. Figure \ref{fig:pyramid1} illustrates the bottom-up approach to simulation validation studies for the composite structure design \cite{JustinHicksFTDM2021}. 
With the detector designs getting increasingly complex an iterative prototype and validation approach is limited to scaled down versions of sub-components or assemblies. However, even small-scale realistic prototypes matched in tolerances and deformation to detailed FEAs help to inform design iterations able to explore many more potential design choices, while saving on cost and time-consuming manufacturing of full scale prototypes. Similarly, the tool shape compensation simulations for manufacturing itself, also benefit from this approach. However, given the different levels of stress between compression molded parts and oven cured carbon fiber laminates this approach is limited to a single production process for layup and autoclave cure or oven-cure of composite pre-preg materials. The current process adopted for the HL-LHC CMS detector upgrade is detailed in Ref.\ \cite{SushrutCPAD2021} and overcomes the challenges for the precision and accuracy in manufacturing of composite support structures. Further progress towards utilizing different composite manufacturing techniques into a single structure rely on an improved manufacturing simulation design cycle - an active research field in composite engineering. 

\subsection{Moving towards multi-functional support structures}

\begin{figure}
    \centering
    \includegraphics[width=\textwidth,height=\textheight,keepaspectratio]{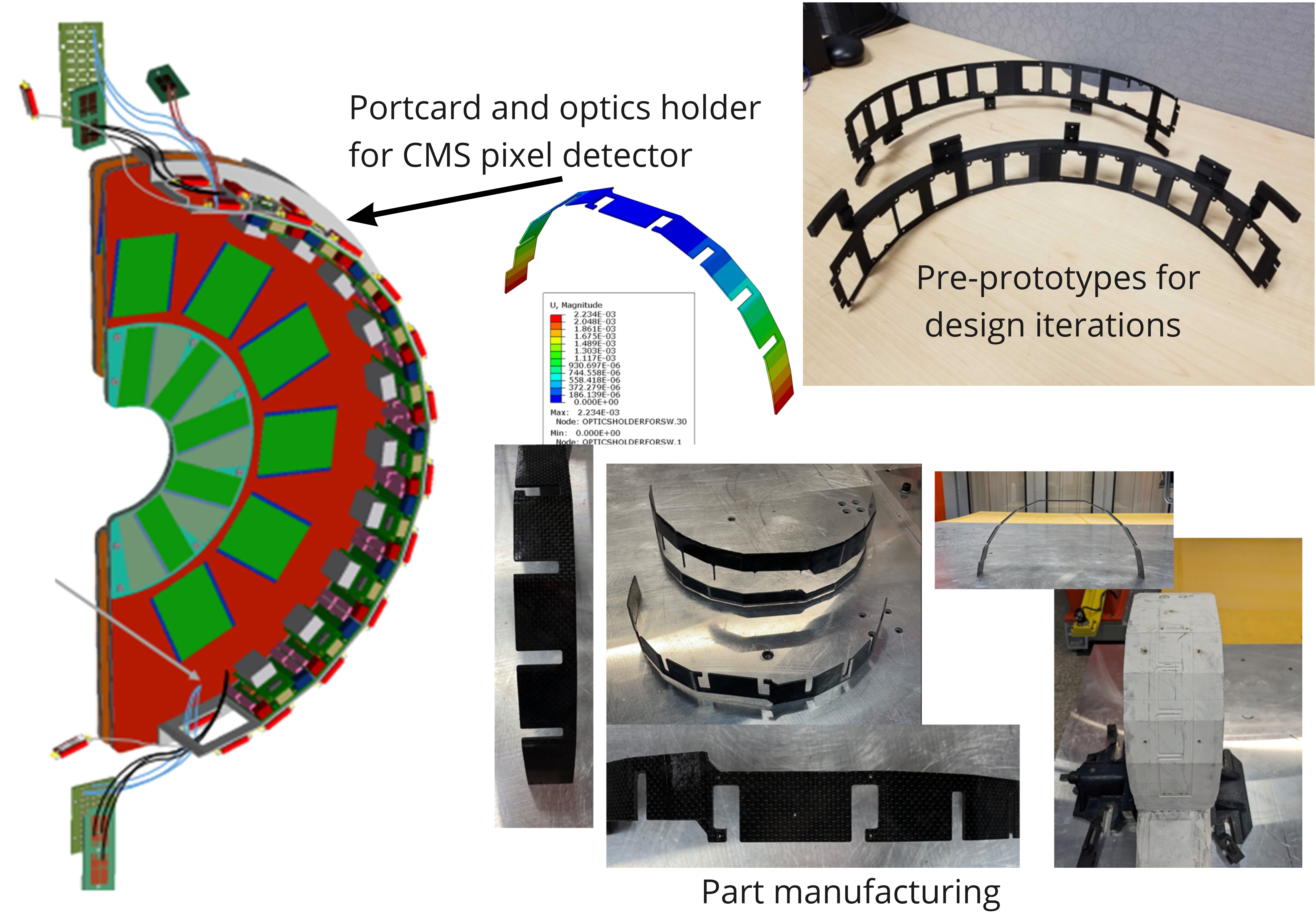}
    \caption{Portcard and optics holders are structural components that are cooled with 2-phase $CO_{2}$ cooling loops. Using highly thermally conductive material for the support structures will enable reducing the mass budget on these components. }
    \label{fig:Portcard}
\end{figure}

The cooling requirements of particle detectors at the MuC or FCC colliders drastically increase compared to those at the HL-LHC environment due to the increase in the collision energy and frequency, as well as corresponding radiation levels. Hence, the mass budget taken by the cooling / thermal management system together with the independent support structure skeleton system imposes bottle necks for the physics performance of the detectors. At Purdue University R\&D efforts are underway to explore multi-functional composite structures, that are manufactured from highly thermally conductive materials. These will create a more optimal thermal pathway between the silicon modules and the cooling loops allowing to further reduce the mass budget of support structures and potentially even reduce the ``density" of cooling loops especially near the interaction point. This reduces the material budget in the immediate vicinity of the interaction point and as a result will enable higher detector performance. The manufacturing process is a combination of additive manufacturing and compression molding with polymer composites filled with additives to increase the thermal conductivity of the composite structure. The added requirement from any detector support structure for maintaining the required stiffness intertwines the necessity of multi-method composite simulations that will establish a workflow to enable composite support structures that are thermally conductive. One such application is a composite structure called ``portcard and optics holder" for the HL-LHC CMS pixel detector. The traditional manufacturing process is illustrated in Figure \ref{fig:Portcard} which has a primary $CO_2$ cooling loop. This additional mass budget can be reduced by using highly thermally conductive polymeric materials that are radiation hard for the structural application. The material selection process starts with studying the manufacturability for materials as well as the radiation hardness. Once the radiation hardness is determined to be adequate for the detector radiation dose estimates, the manufacturing process is studied for the components. A proposed path for multi-method manufacturing consists of 3D-printing the compression molding charges for the portcard holders and then using compression overmolding process with the inserts needed to achieve a multifunctional support structure. The stiffness and thus deformation predictions for such carbon fiber reinforced composites is highly dependent on the orientation of the short and long fibers in the component. Thus the Additive3D simulation\cite{additive3Dbyeduardo, Sushrutsmastersthesis} developed at Purdue University with mold flow simulations for compression overmolding \cite{Bensthesis, Drewswork1}\cite{Drewswork2, favaloro2018simulation} creates a complete toolbox to accomplish that. It also allows to include machine learning based topology optimization to fulfill the low-mass high-stiffness and thermal conductivity structure requirements of future detectors as briefly described in the next Section.

\begin{figure}
    \centering
    \includegraphics[width=\textwidth,height=\textheight,keepaspectratio]{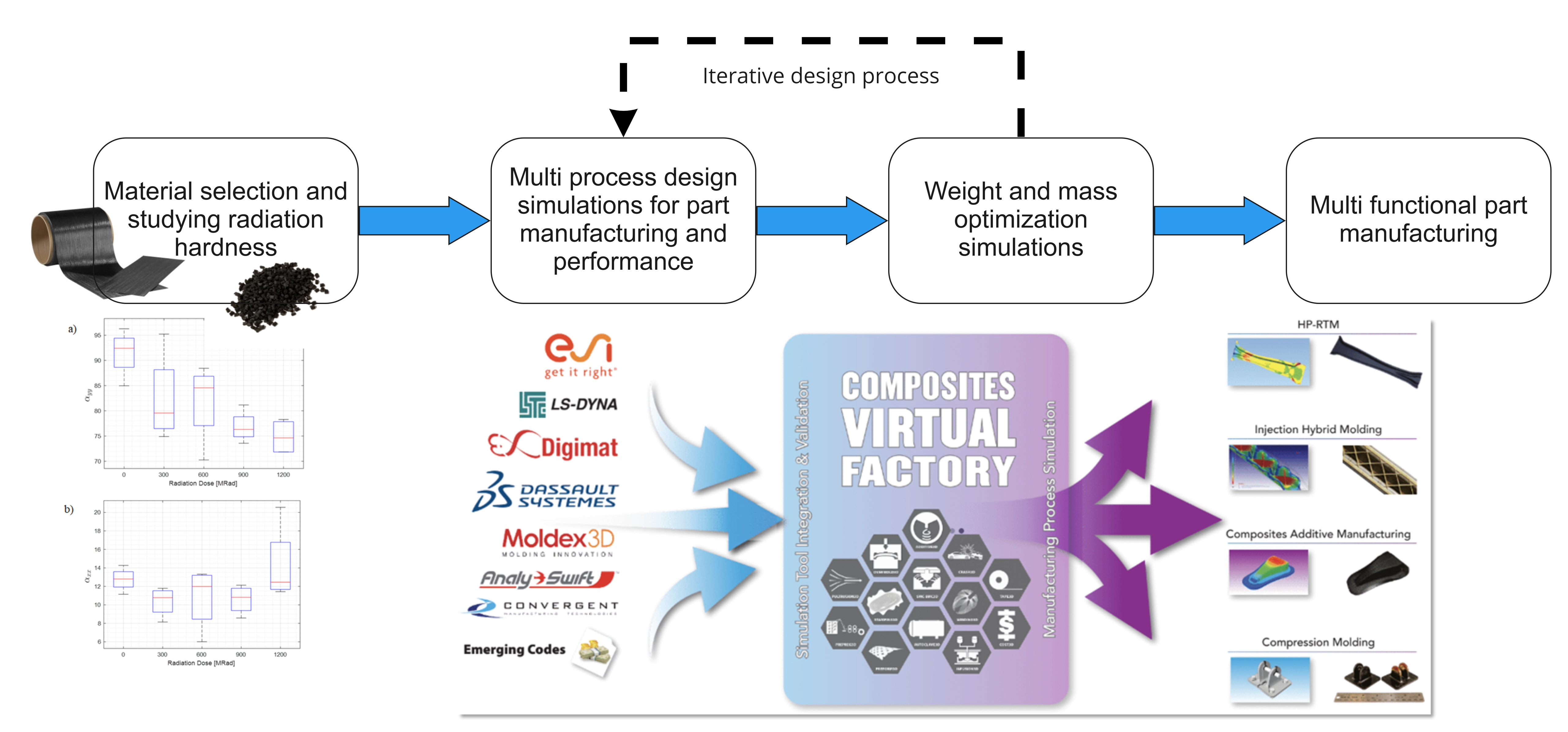}
    \caption{Multi-functional multi-process end to end analysis flow chart for process and performance metrics is presented for future detector mechanics}
    \label{fig:multiprocess_sim}
\end{figure}

\begin{figure}
    \centering
    \includegraphics[width=\textwidth,height=\textheight,keepaspectratio]{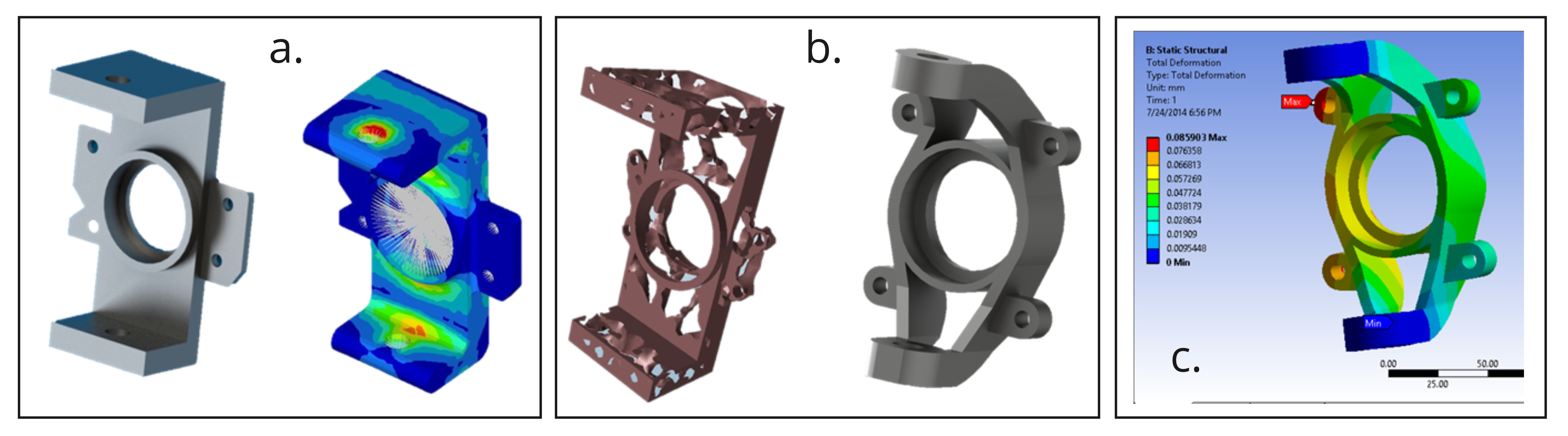}
    \caption{Topology and mass optimization simulation workflow to be integrated with the process and performance simulations developed at Purdue University \cite{bhusari2016fea}. More details in the text.}
    \label{fig:topopti_sim}
\end{figure}

\subsection{Incorporating machine learning based topology and mass optimization}

Topology and mass optimization is an emerging field in the composites design space. The need for light-weight structures for automobile, as well as aerospace applications drive the current topology optimization and mass optimization approaches. Figure \ref{fig:topopti_sim} shows the state-of-the-art approach to mass optimization, where a.\ shows the outcome of a part design being analysed for the loading conditions and boundary conditions. Subfigure b.\ shows the result of iterative machine learning algorithms to obtain the optimal load transfer paths. A final part is designed based on these topology optimization results. Finally, c.\ presents the final design validated with performance simulations prior to any prototyping and part manufacturing thus saving cost and time in deriving a mass-optimized structure. For applications in high energy physics the analysis for topology optimization will be informed by the needs of the thermal pathways to cool the electronics and silicon detectors as well as the mechanical loading requirements for a support structure. The thermal pathways are a bottleneck in most of these structures used in high energy physics detector support mechanics. Machine learning based algorithms will iterate and tie in the optimal thermal network with the minimum structural requirements and boundary conditions for any structure. \\

A large number of future and proposed experiments require silicon trackers including support and service structures with $\le 1\% X_{0}$ (radiation length) per layer, which is extremely challenging while simultaneously increasing segmentation and radiation tolerance, enabling dense jet tracking, improving timing precision, increasing system size, improving manufacturing and replace-ability/maintenance (robotic), and all that with finite research budgets and schedules. Hence, an additional input quantity specific to the design of support structures in high radiation environments will be the radiation length and the consequent minimization of it to reduce as much as possible multiple scattering, which negatively impacts the physics performance. Assuming a support structure manufactured following these design guidelines we can translate the densities of these composite support structure materials into radiation length for a realistic tracking detector at FCC-like machines with coverage up to pseudorapidities\footnote{The pseudorapidity $\eta$ is defined in terms of the polar angle with respect to the colliding beam axis as $\eta = -\ln [\tan (\theta /2)]$.} of 6. The fractional contribution of support structure materials and services grows with higher pseudorapidities and consequently the amount of active and dead material passed through by a charged particle also rises. Material savings due to novel approaches as discussed here have the potential of reduction on the order of 30-50\% depending on phase space and awaiting more detailed R\&D studies. The composite design field is still in its infancy in terms of the manufacturability design of composites using mass and topology optimization as can be seen in \cite{optimizationpaper1, optimizationpaper2, optimizationpaper3}. Thus this is an ideal opportunity to explore the conjunction of latest techniques in composite engineering involving machine learning based algorithms for heat transfer, mechanical loading and micro-to-macro scale material response predictions for the performance of the support structure mechanics of high energy physics detectors.

\section{Conclusions}

We discuss cutting-edge manufacturing techniques that can be applied to the design of particle physics support structures and have the potential to significantly improve the physics performance of future detectors at FCC-like machines or a muon collider. Optimizing the mass of composite structures is possible using machine learning relying on a variety of critical design parameters \& characteristics  relevant to the expected physics performance, among them: thermal pathways to cool the electronics and silicon detectors, mechanical loading requirements for a support structure, as well as radiation length of materials involved. Current R\&D efforts at Purdue University and other institutes provide a headway for future detector support structures relying on transformational novel manufacturing techniques.

\bibliographystyle{JHEP}
\bibliography{snomassMech.bib} 

\providecommand{\href}[2]{#2}\begingroup\raggedright\begin{thebibliography}{10}

\bibitem{Run2TrackerPerformance}
{\scshape CMS} collaboration, \emph{{Strategies and performance of the CMS
  silicon tracker alignment during LHC Run 2}}, {\emph{CMS-TRK-20-001,
  CERN-EP-2021-203} (2021) }.

\bibitem{PhaseII_TDR}
{\scshape CMS} collaboration, \emph{{The Phase-2 Upgrade of the CMS Tracker}},
  {\emph{CERN-LHCC-2017-009, CMS-TDR-014} (2017) }.

\bibitem{PhaseII_TDR_ATLAS}
{\scshape ATLAS} collaboration, \emph{{Technical Design Report for the ATLAS
  Inner Tracker Pixel Detector}}, {\emph{CERN-LHCC-2017-021, ATLAS-TDR-030}
  (2017) }.

\bibitem{Koshy}
A.M.K.~Souvik~Das, Andreas~Jung, \emph{{Thermal studies for the CMS Phase II
  Tracker Forward Pixel detector}},  2019.

\bibitem{msg}
R.~Orzuri, E.~Barocio,  and W.~Yu, \emph{{Experimental and Numerical
  Determination of the Thermal Conductivity Tensor for Composites Manufacturing
  Simulation}}, {\emph{Proceedings of the American Society for Composites}
  {\bfseries 32} (2017) }.

\bibitem{JustinHicksFTDM2021}
J.~Hicks, B.~Denos, G.~Kim, S.R.~Karmarkar, S.~Kulkarni and A.~Jung,
  \emph{{Support Tube Manufacturing Trials, Simulation, and Validation for the
  CMS Inner Tracker Phase II Upgrade}},  2021.

\bibitem{SushrutCPAD2021}
S.~Karmarkar, B.~Denos, J.~Hicks and A.~Jung, \emph{{Design, Simulation,
  Manufacturing, and Validation of Prototypes for the CMS Phase II Tracker
  upgrades}},  2021.

\bibitem{additive3Dbyeduardo}
E.B.~Vaca, \emph{{Fusion bonding of fiber reinforced semi-crystalline polymers
  in extrusion deposition additive manufacturing}}, {\emph{School of Materials
  Engineering, Purdue University West Lafayette, Indiana, USA} (2018) }.

\bibitem{Sushrutsmastersthesis}
S.~Karmarkar, \emph{{Extrusion Deposition Additive Manufacturing for High
  Temperature Tooling}}, Ph.D. thesis, Purdue University, 2018.

\bibitem{Bensthesis}
A.J.~Favaloro, B.R.~Denos, D.E.~Sommer, R.A.~Cutting and J.E.~Goodsell,
  \emph{{Validation of process simulation workflow for thermosetting prepreg
  platelet molding compounds}}, {\emph{Composites Part B: Engineering}
  {\bfseries 224} (2021) 109198}.

\bibitem{Drewswork1}
A.J.~Favaloro, D.E.~Sommer, B.R.~Denos and R.B.~Pipes, \emph{{Simulation of
  prepreg platelet compression molding: Method and orientation validation}},
  {\emph{Journal of Rheology} {\bfseries 62} (2018) 1443}.

\bibitem{Drewswork2}
D.E.~Sommer, S.G.~Kravchenko, B.R.~Denos, A.J.~Favaloro and R.B.~Pipes,
  \emph{{Integrative analysis for prediction of process-induced,
  orientation-dependent tensile properties in a stochastic prepreg platelet
  molded composite}}, {\emph{Composites Part A: Applied Science and
  Manufacturing} {\bfseries 130} (2020) 105759}.

\bibitem{favaloro2018simulation}
A.J.~Favaloro, D.E.~Sommer, B.R.~Denos and R.B.~Pipes, \emph{{Simulation of
  prepreg platelet compression molding: Method and orientation validation}},
  {\emph{Journal of Rheology} {\bfseries 62} (2018) 1443}.

\bibitem{bhusari2016fea}
A.~Bhusari, A.~Chavan and S.~Karmarkar, \emph{{FEA \& optimisation of steering
  knuckle of ATV}},  in \emph{IRF international conference}, 2016.

\bibitem{optimizationpaper1}
J.L.~Pelletier and S.S.~Vel, \emph{{Multi-objective optimization of fiber
  reinforced composite laminates for strength, stiffness and minimal mass}},
  {\emph{Computers \& structures} {\bfseries 84} (2006) 2065}.

\bibitem{optimizationpaper2}
A.E.~Albanesi, I.~Peralta, F.~Bre, B.A.~Storti and V.D.~Fachinotti, \emph{{An
  optimization method based on the evolutionary and topology approaches to
  reduce the mass of composite wind turbine blades}}, {\emph{Structural and
  Multidisciplinary Optimization} {\bfseries 62} (2020) 619}.

\bibitem{optimizationpaper3}
S.~Nikbakt, S.~Kamarian and M.~Shakeri, \emph{{A review on optimization of
  composite structures Part I: Laminated composites}}, {\emph{Composite
  Structures} {\bfseries 195} (2018) 158}.

\end{thebibliography}\endgroup





 


\end{document}